\documentstyle[eqsecnum,preprint,tighten,aps]{revtex}
\newfont{\largemi}{cmmi10}
\newfont{\smallmi}{cmmi6}

\begin{document}
\draft

\title{Generalization of the $N_pN_n$ scheme
to  non-yrast levels of even-even nuclei}
 
\author{ Y. M. Zhao$^{ab}$, 
  and A. Arima$^{c}$}

\vspace{0.2in}
 \address{ $^a$
Department of Physics,
 Southeast University, Nanjing 210018 China \\
$^b$ Department of Physics, Saitama University, Urawa, Saitama 338, Japan\\
$^c$
The House of Councilors, 2-1-1 Nagatacho,  Chiyodaku, Tokyo 100-8962   Japan}

\date{\today}
\maketitle

\begin{abstract}
In this paper we present the systematics of   excitation energies 
for even-even  nuclei in two regions: the 50$<$Z$\le$66, 
82$<$N$\le$104 region,  and
the 66$<$Z$<$82, 82$<$N$\le$104 region.
Using the $N_pN_n$ scheme, we obtain  
 compact trajectories for the ground
 band  as well as quasi-$\beta$ and quasi-$\gamma$ bands.
 This  suggests that the $N_pN_n$ scheme is  useful even if one extends it 
 to  non-yrast levels, and thus can serve  
 as a general tool to disclose new types of structural evolution
  for higher excitations, besides the yrast states 
   which have been investigated extensively.  It is highlighted that
  deformations in non-yrast quasibands of nuclei with $Z \sim 80$ and 
 $N \sim 104$ are often  very different from those in the ground bands.

\end{abstract}


\newpage

The  importance of the residual  valence
p-n interaction 
in the development of collectivity, phase/shape transitions  
and deformation has  been stressed and discussed 
by many authors, such as deShalit and Goldhaber \cite{deShalit}, 
 Talmi \cite{Talmi}, and  Federman and Pittel \cite{Pittel}. 
 This idea was further simplified by Casten  \cite{Casten}: 
Suppose that a  simple product of valence proton number and valence
neutron number, $N_pN_n$,
is a reasonable estimate to gauge this interaction. Then there must be 
some  correlations between the collective observables (such as
excited energies, deformations, etc.) and   $N_pN_n$. 

The $N_pN_n$ plots helped a lot in the past decades for the
classification and a better understanding 
of the increasingly rich of data \cite{Casten0}. However,
almost all  the plots based on the 
$N_pN_n$ scheme were applied to the ground or yrast bands. One curious 
question is whether these simple and naive plots are helpful 
in any ways for the non-yrast 
levels. In this paper we shall find that
the $N_pN_n$ plots are actually 
applicable to  these higher excited states.

The clue lies in a concept--``quasiband",
which was proposed by M. Sakai \cite{Sakai-NPA} in 1967
-- for nearly spherical and/or  transitional nuclei.
It might be inadequate to use the terms such as
$\beta$ or $\gamma$ band(s) for levels of nuclei
near closed shell nuclei,    because 
the physical content  of bands in transitional
and nearly spherical nuclei  
is different from those of deformed nuclei.
Nevertheless,   for the sake of  convenience,
 one may introduce terms such as quasi-ground, 
  quasi-$\beta$ and quasi-$\gamma$ bands  
in spherical and transitional nuclei,   which are regarded
  as the counterparts of collective bands in deformed regions. 
The  quasi-ground bands 
and  quasi-$\beta$ bands thus have   spin sequences 
$0^+$,  $2^+$, $4^+$, $6^+$, etc., and the quasi-$\gamma$ bands
have spin sequences $2^+$, $3^+$, $4^+$, $5^+$, etc. The
data for even-even nuclei using ``quasi"-bands  have been 
compiled and revised many times \cite{Sakai-data}, and have been used 
extensively by many authors. 

The  question relevant to this paper is: when we go from 
 spherical regions with  vibrational motion  to
the transition regions with complicated modes, and finally  to deformed
regions with ground rotational and $\beta$- and $\gamma$-vibrational
bands,  do the  excited  energies of
non-yrast levels evolve smoothly with their $N_pN_n$ values?
Even if the evolution is smooth, those data versus $N_pN_n$ might be very  
scattered and not so useful, 
because the  behaviors  of these
non-yrast levels  are  ${\it not}$ necessarily the
same as those of the yrast band.

It  is therefore  interesting   to 
investigate whether  the excitations based 
on these ``band" heads evolves similarly as the yrast band,
and whether the $N_pN_n$ scheme is relevant to classify
the levels for non-yrast quasi bands, in particular, whether the
$N_pN_n$ scheme is able to highlight some anomalies which
reflect some interesting mechanism in the relevant levels. 

We first look at the situation for the yrast levels. 
We focus on two regions which were studied extensively:
$50 \le Z \le 68$, $84 \le N \le 104$ and 
$68 < Z \le 80$, $84 \le N \le 104$.
For the former region, we use the effective numbers of  
valence protons for nuclei which are affected by the $Z=64$
subshell.  The effective numbers were tabulated  
in a recent Letter by a fit of ground state deformations
using the $N_pN_n$ scheme \cite{Zhao-Letter}.
For the latter region no effective numbers are  used but 
 the nuclei with $Z\sim 78$ were found to exhibit anomalies
of ground-state deformations which are maximized around 
$N=104$ \cite{Zhao-pfactor}, and their $E_{2_1^+}$'s 
are  also abnormal in the $N_pN_n$ plots.  We thus   
discriminate the  $Z > 76$   and   $Z \le 76$ cases  in 
the $68 < Z \le 80$, $84 \le N \le 104$ region: We use
solid squares for  $Z \le 76$ and open squares for 
  $Z > 76$.

All the data used in this paper are taken from Ref. \cite{Sakai-data}. 
Fig. 1 shows the systematics of $E_{2_1^+}$ versus $N_pN_n$.  
 We see that there exists a very strong correlation
between the $E_{2_1^+}$'s and the $N_nN_p$ values, if 
we reasonably offset the subshell effect and 
exclude the anomalies of  $Z \sim 78$ and $N\sim 104$.

For non-yrast states, we define $E'_{3^+}(\gamma_1) = 
E_{3^+}(\gamma_1) - E_{2^+}(\gamma_1)$, and 
 $E'_{2^+}(\beta_1) =  E_{2^+}(\beta_1) - E_{0^+}(\beta_1)$.
It is noted that some authors would not use
the above appellation of ``$\beta_1$ band" (which may imply 
a very specific structure that  may not be true in $any$
given nucleus), but would use the term of the ``lowest excited $K$=0 
band". However,  we keep to Sakai's notation for short  in this paper. 
The behavior of these $E'$ values is then assumed to
reflect the structural  evolution of the non-yrast states. 

Fig. 2 plots $E'_{3^+}(\gamma_1)$  and $E'_{2^+}(\beta_1)$ versus  
their $N_pN_n$ values.
Here  the same $N_nN_p$ numbers
as in the Fig. 1 are used. 
Despite of the fact that the behaviors of excited states
in the ground band may be different from  those
in non-yrast bands (say, the $\beta_1$ or $\gamma_1$ band here),
very interestingly, the $N_nN_p$ scheme continues to work very well. 
A strong correlation between the $E'_{3^+}(\gamma_1)$ or $E'_{2^+}(\beta_1)$
values  and their corresponding $N_nN_p$ is easily seen.

The above  $N_pN_n$ plots not only provide  a
naive correlation between the excited energies
of non-yrast levels and  their  $N_pN_n$ values, but 
also provide  more insights into their structural evolution. 
For the ground band,    nuclei with
$Z=78$ and 80 are much more deformed than they ``should" be, with 
$N\sim 104$ nuclei deviating furthest from the correlation,
which was shown in \cite{Zhao-pfactor}. 
Fig. 1 (b) highlights the same anomalies in their  
$2_1^+$ energy levels. 
Therefore, one would easily have the intuition  that the
non-yrast   levels of  these nuclei deviate
from the correlation  in a similar
manner.  However, Fig. 2(d) shows that the  $E'_{3^+}(\gamma_1)$ 
values of $Z=78$ and $N=102$, 104 are quite ``normal".
If it is assumed  that  a large
deformation is associated with 
  smaller excitation energies within a band, as is well known in the
  ground band, the present observation indicates 
a very drastic change in deformation when we go from the ground band to  
the $\gamma_1$ band for the nuclei $_{78}^{180}$Pt$_{102}$
and $_{78}^{182}$Pt$_{104}$.     It would be very interesting to check 
 the  data for $E'_{3^+}(\gamma_1)$ for the  Hg isotopes
 with $N\le 104$ when  they will be available in the future. 
The  deformation of $_{80}^{184}$Hg$_{104}$
in  the $2^+ (\beta_1)$ state in the  $\beta_1$ band, 
as discussed above, is suggested
to be  quite close to that
of the ground state, while the deformation of  
 $_{78}^{182}$Pt$_{104}$ in the $2^+ (\beta_1)$ state  
  is  substantially smaller  than
  in the ground state. 
Therefore, based on  the $N_pN_n$ plots of this paper  and
assuming a correlation between the values of deformation and excited 
energies in the quasi-bands discussed in \cite{Sakai-NPA,Sakai-data},
we suggest that the  deformations of non-yrast levels of 
 nuclei with $N \sim 104$ and $Z\sim 78$ are  
 complicated, and very state- or band-dependent even in the
  low-lying and low-spin  states.

To summarize, compact trajectories of excitation energies of 
both the ground band and the quasi-$\beta$ and quasi-$\gamma$ bands 
for  even-even  nuclei 
have been obtained in the  $N_pN_n$  plots, showing that
the   $N_pN_n$ scheme is also useful for non-yrast states, 
despite of the fact that the behavior of the
quasi-$\beta$ and quasi-$\gamma$ bands might be different from that
of the yrast one.  These plots 
highlight the anomalies of both the ground band and higher
excited bands. We suggest,  using this 
simple scheme, that 
 the deformations of non-yrast (but very low) levels of 
  nuclei with $N \sim 104$ and $Z\sim 78$ are  
  very  state- or band-dependent. 
   The $N_pN_n$ scheme  is thus very useful 
 to disclose anomalies  of   structural evolution, especially
 for those nuclei without rich data.

The authors would like to thank Professors Rick F. Casten and
 Stuart Pittel for their discussions and  constructive comments. 
 This work is supported in part by
 the Japan Society of Promotion of Science (Contract No. P01021). 

\newpage

Captions:

\vspace{0.4in}

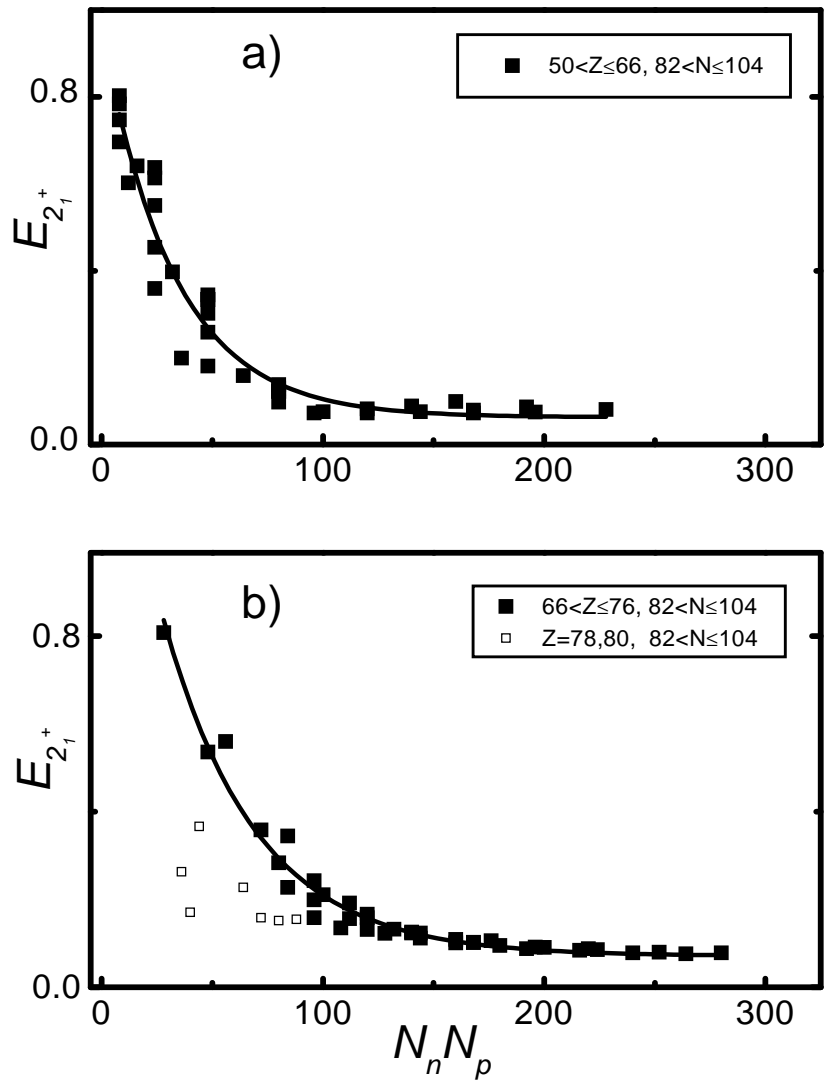
{FIG. 1. The  $E_{2_1^+}$ versus  $N_pN_n$  for nuclei with 
a) 50$<$Z$\le$66 and 82$<$N$\le$104;
b) 66$<$Z$<$82 and 82$<$N$\le$104.
In a) effective numbers of valence protons are taken from \cite{Zhao-Letter}.
In b) the nuclei with $Z=78$ and 80 are labeled with open squares. 
 See the text for details.  }

\vspace{0.4in}

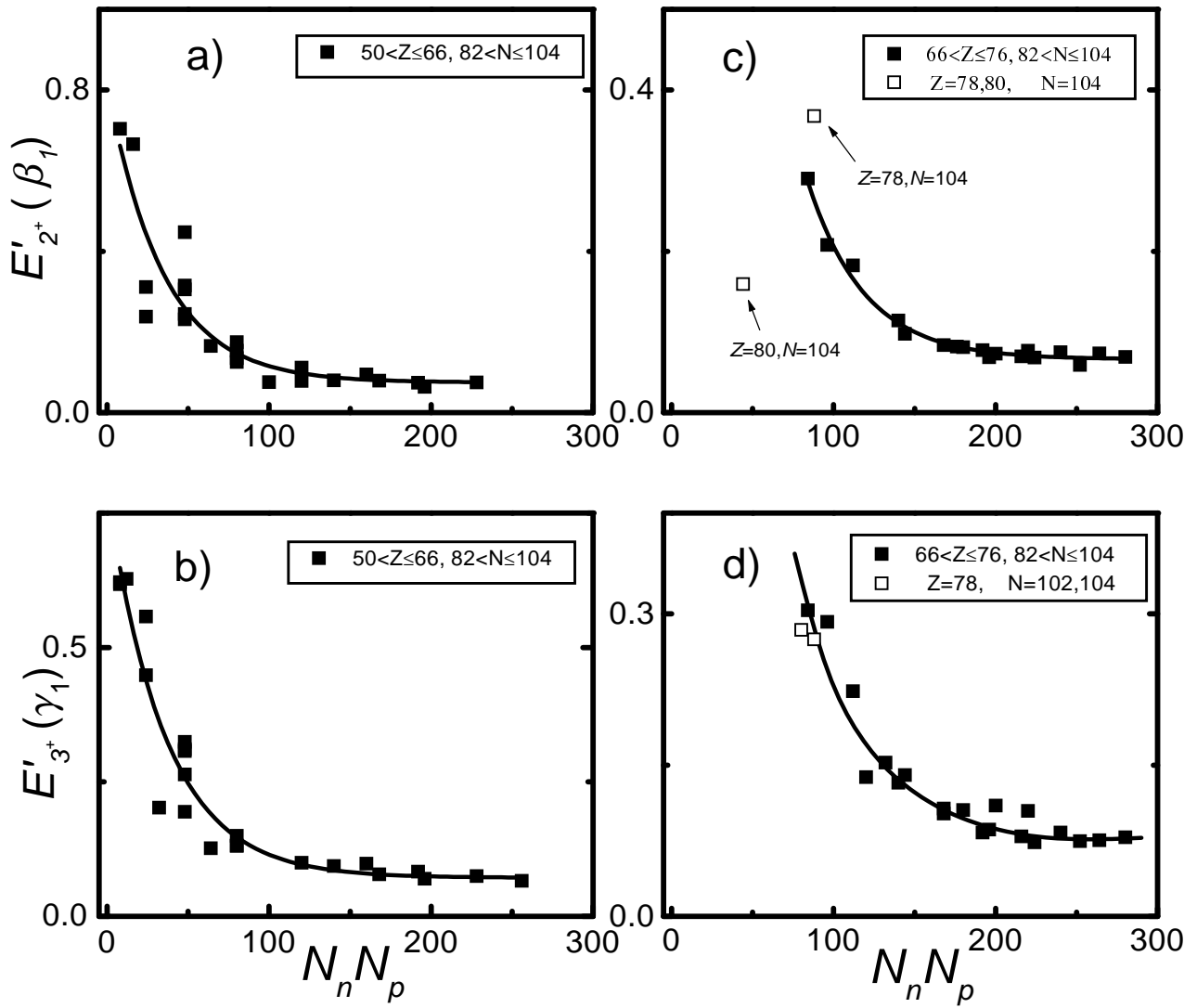
{FIG. 2.  The excited energy values of quasi-$\beta$ and
quasi-$\gamma$ bands  versus  $N_pN_n$. In a) and b)
$50 < Z \le 68$, $84 \le N \le 104$. In c) and d) 
$68 < Z \le 80$, $84 \le N \le 104$.
The  $E'_{2_1^+} (\beta_1)$ are given in a) and c)
and  $E'_{3_1^+} (\gamma_1)$ are given in b) and d), respectively. }

\end{document}